# Anomalous quantum Griffiths singularity in ultrathin crystalline lead films

Yi Liu[1,2,†], Ziqiao Wang[1,2,†], Pujia Shan[1,2], Chaofei Liu[1,2], Yue Tang[1,2], Cheng Chen[1,2], Ying Xing[1,2], Qingyan Wang[1,2], Haiwen Liu[3,*], Xi Lin[1,2], X. C. Xie[1,2], Jian Wang[1,2,4,*]

[1]*International Center for Quantum Materials, School of Physics, Peking University, Beijing 100871, China.*

[2]*Collaborative Innovation Center of Quantum Matter, Beijing 100871, China.*

[3]*Center for Advanced Quantum Studies, Department of Physics, Beijing Normal University, Beijing 100875, China.*

[4]*CAS Center for Excellence in Topological Quantum Computation, University of Chinese Academy of Sciences, Beijing 100190, China.*

†These authors contributed equally to this work.

*Corresponding authors: Haiwen Liu (haiwen.liu@bnu.edu.cn) and Jian Wang (jianwangphysics@pku.edu.cn)

**We study the magnetic field induced superconductor-metal transition (SMT) in ultrathin crystalline Pb films. By performing ultralow temperature transport measurement, the divergent critical exponent as an indication of quantum Griffiths singularity (QGS) is observed when approaching zero temperature quantum critical point. Distinctively, the phase boundary of SMT exhibits an anomalous behavior in low temperature regime that the onset critical field decreases with decreasing temperatures, which distinguishes our observation from earlier reports of QGS in various two-dimensional superconductors. We demonstrate that this observed anomalous phase boundary has its origin from the superconducting fluctuations in ultrathin Pb films. Our findings reveal a novel aspect of the QGS of SMT in two-dimentional superconducting systems with anomalous phase boundary.**

As a paradigm of quantum phase transition (QPT) [1,2], superconductor-insulator/metal transition (SIT/SMT) has been widely investigated in two-dimensional (2D) superconductors over the past thirty years [3-11]. When the external magnetic field slightly surpasses the critical value $B_c$ of the quantum critical point, superconductivity will be destroyed and the system enters an insulating state or a weakly localized metal state. In general, a conventional Bardeen-Cooper-Schrieffer (BCS)-type superconductor has a clear phase boundary with a monotonically negative correlation between the upper critical field and temperature. However, in high temperature and organic superconductors, as well as disordered low-dimensional superconducting systems, the relatively low superfluid density leads to predominant fluctuation effect, and as a consequence the phase boundary between superconducting state and normal state becomes indistinct [12-14]. Previous experimental work has shown that the superfluid density of ultrathin Pb films decreases dramatically from the bulk value [15], indicating a significantly larger role of fluctuation in Pb films compared to the bulk counterpart. Moreover, superconducting fluctuation leads to the formation of Cooper pairs in the normal state, which largely influences the characteristic of the phase boundary [16,17]. For instance, the superconducting fluctuation can give rise to a reentrant behavior [18,19], where the sheet resistance of the system firstly drops with decreasing temperatures and then rises at lower temperatures. The special reentrant behavior would result in a complicated and anomalous phase boundary, which has not been fully understood and requires further experimental investigations and theoretical analysis.

Recently, crystalline 2D superconductors offer a new perspective to explore unprecedented quantum phenomena [20], such as the observation of Zeeman-protected Ising superconductivity [21-25] and the emergence of quantum metal state in highly crystalline 2D systems [26-29]. Among them, one striking phenomenon is the quantum Griffiths singularity (QGS) of SMT characterized as a divergent critical exponent $zv$ at zero temperature quantum critical point due to the formation of large rare regions [24,29-34]. QGS has been reported in a wide range of crystalline 2D superconductors, such as 3-monolayer (ML) Ga film [32], 1-ML $NbSe_2$ film [24], $LaAlO_3/SrTiO_3$(110) interface [34], gated $MoS_2$ and ZrNCl [29]. These observations of QGS indicate that dissipation and quenched disorder have dramatic effect on the SMT. In the ultralow temperature regime, the quenched disorder leads to large local superconducting islands (rare regions), and the size of these islands increase exponentially when approaching zero temperature [31]. The slow dynamics (such as the relaxation originating from the lowest energy level) of these large superconducting islands give rise to a divergent critical exponent $zv$ in the SMT of low-dimensional superconductors [30,31]. However, the fate of QGS in 2D systems with strong superconducting fluctuations remains an open question. Thus, researches on QGS of SMT under the influence of strong superconducting fluctuation in a crystalline 2D superconductor are highly desired.

In this paper, we report a novel type of QGS in ultrathin crystalline Pb film, which exhibits an anomalous phase boundary of SMT in low temperature regime. The macro-size atomically flat Pb films were epitaxially grown on striped incommensurate phase on Si(111) substrate in an ultrahigh-vacuum molecular beam epitaxy (MBE) chamber. By systematic transport measurement at ultralow temperatures, the ultrathin Pb film undergoes a magnetic field induced SMT. Scaling analysis shows that the critical exponent of SMT diverges when approaching quantum critical point, as an indication of QGS. However, the onset critical magnetic fields of Pb film determined by the crossing points of magnetoresistance isotherms decrease significantly with decreasing temperature in low temperature regime, which shows pronounced differences from normal QGS. Further theoretical analysis reveals that the anomalous phase boundary of SMT can be quantitatively explained by the superconducting fluctuation in ultrathin Pb films. This anomalous phase boundary leads to the reentrant behavior of $R_s$(T) curves, and scaling analysis of the SMT within the reentrant region reveals a new type of quantum critical point with anomalous QGS.

Figure 1 presents the superconducting properties of 4-ML Pb film down to 0.5 K measured in a commercial Physical Property Measurement System (Quantum Design, PPMS-16) with the Helium-3 option. The schematic of the standard four-electrode transport measurement is shown in the inset of Fig. 1(a). The superconducting transition occurs at $T_c^{onset} = 9.21$ K indicated by the deviation from linear extrapolation of the normal state resistance $R_n$, which is even higher than the $T_c$ of bulk Pb ($T_c = 7.2$ K). With decreasing temperature, the sheet resistance $R_s$ drops to zero within the measurement resolution at $T_c^{zero} = 5.73$ K [Fig. 1(a)]. Figure 1(b) reveals the $R_s(T)$ curves measured at different perpendicular magnetic fields. Distinct from the usual SMT with monotonic phase boundary separating the regions of $dR/dT < 0$ and $dR/dT > 0$, $R_s(T)$ curves in 4-ML Pb films exhibit a pronounced reentrant behavior at lower temperatures [Fig. 1(b)]. Specifically, when applying magnetic field slightly over 3.3 T, the sheet resistance firstly decreases with decreasing temperature, reaches the minimum at $T_{min}$ and then rises at lower temperature. This reentrant behavior, including the fact that $T_{min}$ increases with increasing magnetic field, can be qualitatively reproduced after taking the superconducting fluctuation into account [16,17,35].

To systematically investigate the anomalous SMT behavior, we measured the temperature-dependent magnetoresistance in details. Interestingly, the $R_s(B)$ curves cross each other in a relatively large transition region instead of a critical point [Fig. 2(a)]. The crossing points of $R_s(B)$ curves at neighboring temperatures are summarized in Fig. 2(b) as black circles, which show a large transition region from 3.29 T to 4.15 T. We also plot the field dependent onset critical temperature $T_c^{\text{onset}}(B)$ in the same figure as blue circles to show the phase boundary at higher temperatures. Here, $T_c^{\text{onset}}(B)$ is defined as the temperature where the $R_s(T)$ curve begins to deviate from the linear extrapolation of the normal state resistance. The measured SMT phase boundary in 4-ML Pb film is anomalous since it bends down at low temperatures below 2.44 K. The special phase boundary is quantitatively consistent with the theoretical expectation (orange circles) from the plateaus ($dR/dT = 0$) of theoretical $R_s(T)$ curves in Fig. S1 [36], confirming that the anomalous SMT behavior can be ascribed to the superconducting fluctuation in 4-ML Pb film.

Following the previous work [1,2], we use scaling analysis to determine the critical behavior of the anomalous SMT observed in 4-ML Pb film. $R_s(B)$ curves at neighboring temperatures are classified into one group so that the critical transition region can be approximately regarded as a single "critical" point ($B_c, R_c$) [Fig. 2(c), Fig. S2 and S3]. Thus we can apply the standard finite size scaling in each approximate crossing point to get the effective "critical" parameters. Generally, the scaling dependence of sheet resistance on temperature and magnetic field takes the form [3,32]: $R_s(B, T) = R_c \cdot F(|B - B_c|T^{-1/z\nu})$, where $F$ is an arbitrary function with $F(0) = 1$, $z$ and $\nu$ are the dynamical critical exponent and correlation length exponent, respectively. We plot the scaling curves of $R_s(B)/R_c$ at various temperatures against the scaling variable $|B - B_c|t$, where $t = (T/T_0)^{-1/z\nu}$ and $T_0$ is the lowest temperature of the group. Here, the parameter $t$ at each temperature $T$ is determined by performing a rescale of scaling variable $|B - B_c|t$ to make the isotherms $R_s(B)/R_c$ at $T$ match the curve at lowest temperature $T_0$ [Fig. S2 and S3]. The effective "critical" exponent $z\nu$ is acquired by the linear fitting between $\ln T$ and $\ln t$. For instance, scaling analysis on $R_s(B)$ curves from 500 mK to 700 mK yields $z\nu = 3.42 \pm 0.25$ [Fig. S2(b)]. As shown in Fig. 2(c), $z\nu$ grows rapidly and seems to diverge with the temperature decreasing towards zero and the magnetic field tending to a certain critical value $B_c^*$. The divergent behavior of $z\nu$ can be well described by an activated scaling law $z\nu \propto |B - B_c^*|^{-0.6}$ [37,38], indicating the existence of QGS with the infinite-randomness quantum critical point (IRQCP) at $B_c^*$ [Fig 2(c)]. It is noteworthy to mention that, as a result of the anomalous phase boundary, the critical exponent $z\nu$ of anomalous QGS diverges with decreasing field, which is opposite to normal QGS. This uniqueness distinguishes the critical behavior in ultrathin Pb film from previous observations of normal QGS in 2D superconducting systems, including Ga trilayers [32], NbSe$_2$ monolayer [24], LaAlO$_3$/SrTiO$_3$(110) interface [34], gated MoS$_2$ and ZrNCl [29].

To confirm the anomalous QGS in ultralow temperature regime, we measured 4-ML sample down to 60 mK in a dilution refrigerator MNK 126-450 system (Leiden Cryogenics BV). Figure 2(d) shows the $R_s(B)$ curves from 60 mK to 2.30 K and their crossing points are summarized in Fig. 2(e). The temperature dependence of critical field $B_c$ in low temperature region is consistent with the data from PPMS measurement. As shown in Fig. 2(f), the divergence of critical exponent $z\nu$ can be well fitted using the activated scaling law $z\nu \propto |B - B_c^*|^{-0.6}$ down to 60 mK, providing solid evidence of anomalous QGS in ultrathin Pb film.

We then investigate the anomalous QGS in 3.5-ML Pb films. Figure 3 displays the transport properties of 3.5-ML Pb film under perpendicular magnetic field, which are generally similar to the phenomena observed in 4-ML Pb films. Specifically, $R_s(T)$ curves at various fields have a reentrant behavior at low temperatures [Fig. 3(a)], and $R_s(B)$ curves at different temperatures reveal a large transition region [Fig. 3(b)]. The SMT phase boundary, which is determined by the crossing points of $R_s(B)$ curves, exhibits an anomalous decrease of critical field $B_c$ with decreasing temperature in low temperature regime [Fig. 3(c)]. Moreover, the critical exponent $z\nu$ also shows a divergent behavior $z\nu \propto |B - B_c^*|^{-0.6}$ when approaching the quantum critical point with $B_c^* = 2.852$ T [Fig. 3(d)]. In the combination of these observations, we confirm the anomalous QGS in 3.5-ML Pb film. As shown in Fig. S5 [36], scanning tunneling microscopy (STM) image indicates that the surface disorder of 3.5-ML Pb film is much more pronounced than that of 4-ML Pb film. However, the surface disorder in nanometer scale has little effect on the activated scaling law around the anomalous QGS [Fig. 2 and Fig. 3].

In the ultrathin Pb films, the pronounced superconducting fluctuation effect largely changes the shape of the phase boundary. Larkin and his collaborators propose that superconducting fluctuation enhances the conductivity due to the contribution of fluctuating Cooper pairs [16,17]. On the other hand, the formation of Cooper pair also reduces the density of states (DOS) of quasiparticle and decreases the conductivity [12]. The combinational effect of the aforementioned two mechanisms gives rise to the buckle shape of phase boundary around the quantum critical point as shown in Fig. 4. In specific, when decreasing temperature with magnetic field locating in the regime $[B_C^*, B_C']$, the Aslamazov-Larkin type superconducting fluctuation enhances the conductivity, while the influence of superconducting fluctuation on the DOS reduces the conductivity. Comparing with previous systems with normal phase boundary [24,29,32,34], the pronounced fluctuation effect in the ultrathin Pb films could be attributed to the relatively low mobility. Thus, in ultrathin Pb films, the superconducting fluctuation gives rise to remarkable influence on the conductivity of the system, and results in the anomalous phase boundary shown in Fig. 4. When approaching the quantum critical point along this anomalous phase boundary, rare regions of large superconducting islands emerge and dominate the dynamical property of system. The scaling analysis demonstrates that the activated scaling law $z\nu \propto |B - B_C^*|^{-0.6}$ still holds along this anomalous phase boundary, and supports the universality of QGS around the IRQCP in SMT. Our work indicates that, under the influence of the superconducting fluctuation, the anomalous QGS may exist in various superconducting systems with the reentrant behavior.

Other mechanisms can also give rise to an anomalous phase boundary. Firstly, a strong spin-orbit interaction (SOI) will bend down the critical field at low temperatures in Werthamer-Helfand-Hohenberg(WHH) theory [39]. However, the WHH with SOI cannot quantitatively reproduce the anomalous phase boundary in our measurements. The WHH framework also cannot explain the reentrant phenomena, which is usually attributed to fluctuation effect [16,17]. Secondly, the competition between antiferromagnetism and superconductivity can also give rise to the reentrant behavior [40]. But this mechanism is incompatible with our system, since there has not been any report of antiferromagnetism in crystalline Pb thin film. Lastly, in previous studies of amorphous superconducting films, the reentrant behavior is attributed to the efficient Josephson coupling at high temperature and the Coloumb blockade effect at low temperatures [41]. However, this process mainly occurs in amorphous or granular systems with large normal resistance, thus ceases to exist in the crystalline Pb thin films.

In summary, the magnetic field induced SMT is systematically studied in ultrathin crystalline Pb films by *ex situ* transport measurements. Surprisingly, the phase boundary of SMT exhibits an anomalous behavior in low temperature regime that the onset critical field decreases with decreasing temperatures, which can be quantitatively explained by the effect of superconducting fluctuation. Furthermore, the critical exponent of SMT diverges when approaching zero temperature quantum critical point, indicating the existence of anomalous QGS of SMT in 2D superconducting system showing reentrant superconducting behaviors.

We thank Hailong Fu, Pengjie Wang for the help in ultralow temperature transport measurement. This work was financially supported by the National Key Research and Development Program of China (Grant No. 2018YFA0305604, No. 2017YFA0303300, No. 2015CB921102 and No. 2017YFA0304600), the National Natural Science Foundation of China (Grant No.11774008, No. 11534001 and No. 11674028), and the Strategic Priority Research Program of Chinese Academy of Sciences (Grant No. XDB28000000). H. -W. Liu also acknowledges support from the Fundamental Research Funds for the Central Universities.

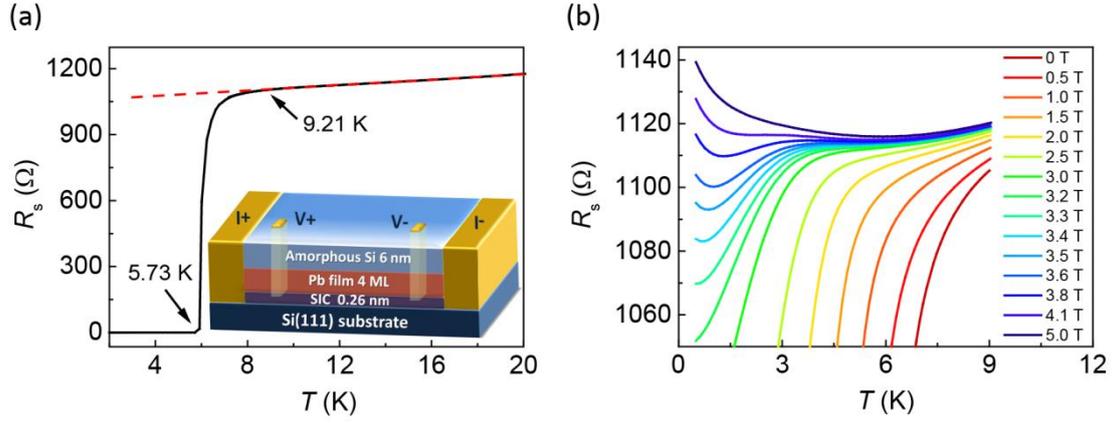

FIG. 1 Transport properties of 4-ML Pb film. (a), Temperature dependence of sheet resistance $R_s$ at zero magnetic field, showing $T_c^{onset}$ = 9.21 K and $T_c^{zero}$ = 5.73 K. The inset is a schematic for standard four-electrode transport measurements. (b), $R_s(T)$ curves measured under various perpendicular magnetic fields up to 5.0 T, revealing clear superconductor-metal transition and reentrant behavior.

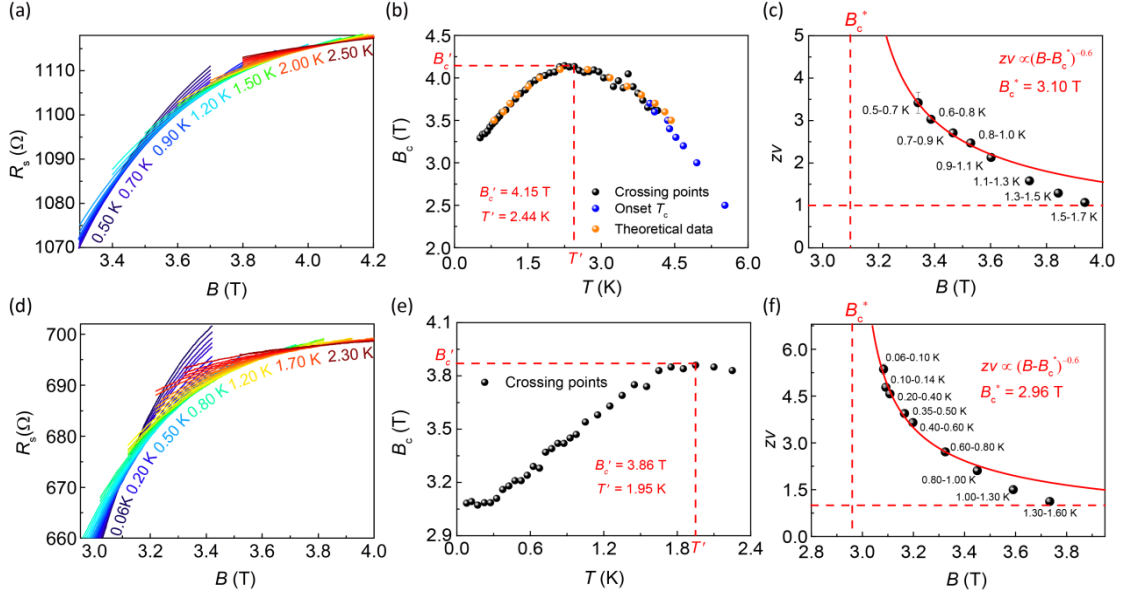

FIG. 2 Anomalous quantum Griffiths singularity of SMT in 4-ML Pb films. (a) $R_s(B)$ curves measured in PPMS from 0.5 K to 2.5 K, revealing a broad crossing region. (b) The crossing points from neighboring $R_s(B)$ curves (black circles) and onset $T_c$ from $R_s(T)$ curves (blue circles) determine the phase boundary of SMT, which is quantitatively consistent with the superconducting fluctuation theory (orange circles) [35]. (c) Scaling analysis indicates anomalous quantum Griffiths singularity behavior in 4-ML Pb film. The critical exponent $z\nu$ diverges as $z\nu \propto |B - B_c^*|^{-0.6}$ when temperature approaches zero Kelvin and magnetic field approaches $B_c^*$. (d)-(f) Similar to (a)-(c), but the data was measured down to 60 mK in a dilution refrigerator, exhibiting consistent anomalous quantum Griffiths singularity behavior.

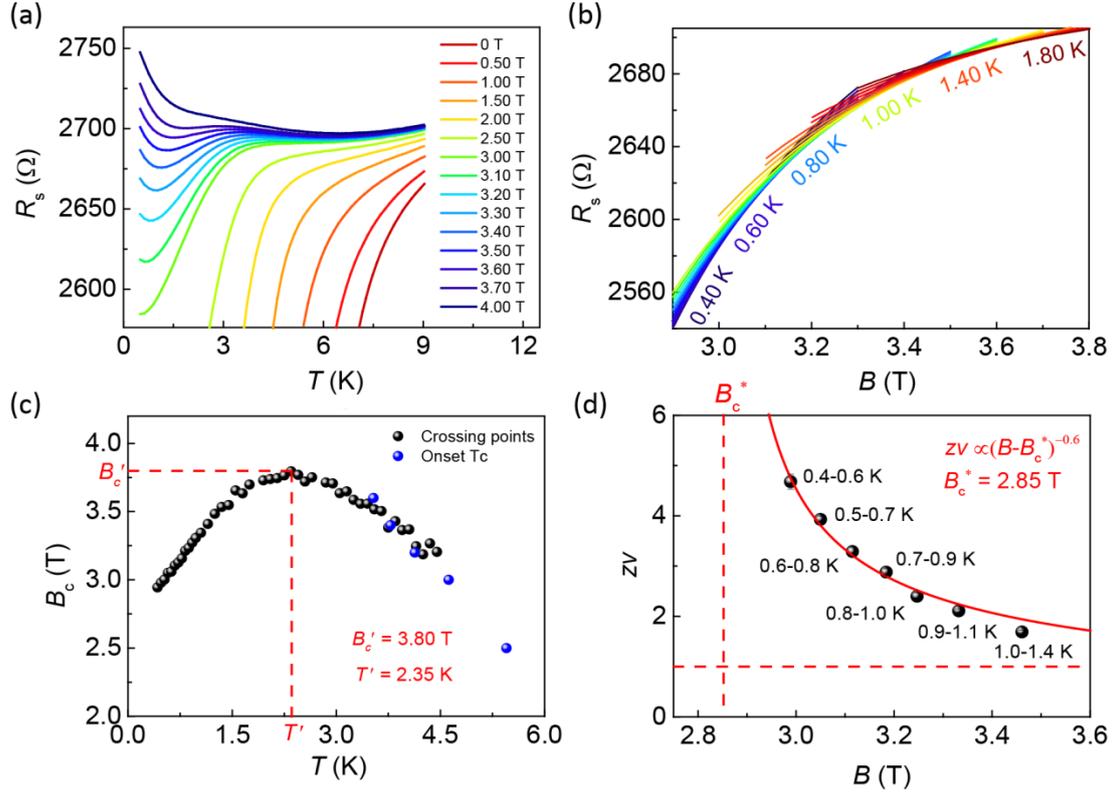

FIG. 3 Anomalous quantum Griffiths singularity of SMT in 3.5-ML Pb film. (a) $R_s(T)$ curves at various magnetic fields show SMT and reentrant behavior. (b) $R_s(B)$ curves measured in the temperature region from 0.4 K to 1.8 K. (c) The crossing points from neighboring $R_s(B)$ curves and the onset $T_c$ from $R_s(T)$ curves determine the phase boundary of SMT. The phase boundary abnormally bends down at lower temperatures. (d) Scaling analysis indicates anomalous QGS behavior in 3.5-ML Pb film. The critical exponent diverges as $zv \propto |B - B_c^*|^{-0.6}$ with temperature approaching zero and magnetic field tending to $B_c^*$.

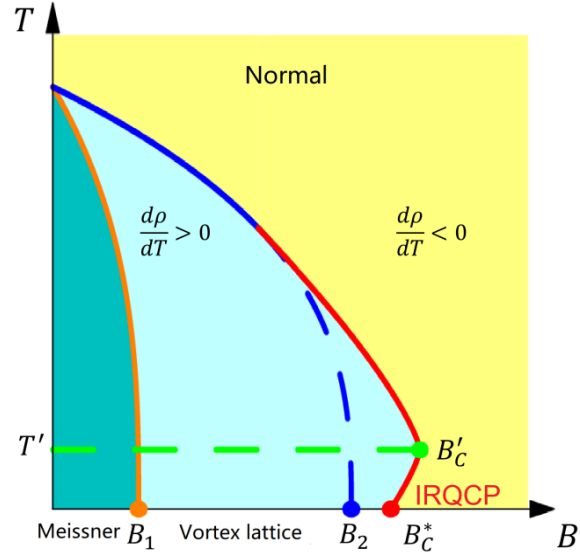

FIG. 4 Schematic of the B-T phase diagram of 2D SMT with pronounced fluctuation effect. $B_2$ is the value of upper critical field with mean field theory. Under the influence of superconducting fluctuations, the mean field phase boundary (blue dashed line) buckles outward to the solid red line, which gives rise to the reentrant phenomena shown in Fig. 1(b) and 3(a). The anomalous phase boundary exists in the regime $[B_c^*, B_c']$. When approaching the real IRQCP $B_c^*$ along this anomalous phase boundary, the system exhibits anomalous QGS behavior.

# Supplemental Material

# Anomalous quantum Griffiths singularity in ultrathin crystalline lead films

Yi Liu[1,2,†], Ziqiao Wang[1,2,†], Pujia Shan[1,2], Chaofei Liu[1,2], Yue Tang[1,2], Cheng Chen[1,2], Ying Xing[1,2], Qingyan Wang[1,2], Haiwen Liu[3,*], Xi Lin[1,2], X. C. Xie[1,2], Jian Wang[1,2,4,*]

[1]*International Center for Quantum Materials, School of Physics, Peking University, Beijing 100871, China.*

[2]*Collaborative Innovation Center of Quantum Matter, Beijing 100871, China.*

[3]*Center for Advanced Quantum Studies, Department of Physics, Beijing Normal University, Beijing 100875, China.*

[4]*CAS Center for Excellence in Topological Quantum Computation, University of Chinese Academy of Sciences, Beijing 100190, China.*

†These authors contributed equally to this work.

*Corresponding authors: Haiwen Liu (haiwen.liu@bnu.edu.cn) and Jian Wang (jianwangphysics@pku.edu.cn)

**Contents**

I. Sample growth and transport measurement

II. Theory of Superconducting fluctuation

III. Figures and Tables

## I. Sample growth and transport measurement

The ultrathin crystalline Pb (111) films were grown in an Omicron ultrahigh vacuum molecular beam epitaxy (MBE) chamber at a base pressure lower than $1 \times 10^{-10}$ mbar. The Si(111) – 7×7 reconstruction phase was prepared by flashing Si(111) substrate at $T \sim 1400$ K for 5-10 times. Before film growth, the striped incommensurate (SIC) Pb phase was acquired by depositing 1.5 ML Pb from a Knudsen cell at room temperature and then annealing at $T \sim 573$ K for 30 sec. The ultrathin Pb (111) films were then grown by depositing Pb atoms on SIC phase at 150 K with a growth rate of ~ 0.2 ML/min and subsequent annealing at room temperature for several minutes. Film growth was monitored by reflection high-energy electron diffraction (RHEED) and characterized by scanning tunneling microscopy (STM).

For *ex situ* transport measurements, Pb films were protected by depositing 6-nm thick amorphous Si capping layer before exposure to the atmosphere. The resistance and magnetoresistance were measured using the standard four-probe method in a commercial Physical Property Measurement System (Quantum Design, PPMS-16) with the Helium-3 option for temperatures down to 0.5 K under perpendicular magnetic field. The ultralow temperature experiment was carried out in a Dilution Refrigerator MNK 126-450 system (Leiden Cryogenics BV) down to 60 mK. Standard low-frequency lock-in technique was used during the measurements with a current excitation of 50 nA at 23Hz for ultra-low temperature measurements. It is noteworthy to mention that the ultralow temperature measurements (Fig. 2(d)-(f)) were carried out right after the growth of sample, so the exposure time to atmosphere is shorter and the sheet resistance is smaller compared to the data from PPMS (Fig. 1 and Fig. 2(a)-(c)).

## II. Theory of Superconducting fluctuation

For two-dimensional superconductors in the dirty limit, the superconducting fluctuation correction to the conductivity is given by Larkin and his collaborators [1,2]. The superconducting fluctuation induced conductivity correction can be written in the form [1, 2]:

$$\delta\sigma = \frac{e^2}{\pi^2 \hbar}[\alpha I_\alpha(b,t) + \beta I_\beta(b,t)] \quad (1)$$

with

$$I_\alpha(b,t) = \ln\frac{r}{b} - \frac{1}{2r} - \psi(r) \quad (2)$$

and

$$I_\beta(b,t) = r\psi'(r) - \frac{1}{2r} - 1, \quad (3)$$

where $r = \frac{b}{3.562t}$, $t = T/T_c \ll 1$, $b = [B - B_{c2}(T)]/B_{c2}(0) \ll 1$, $\psi(r)$ is the digamma function and $B_{c2}(T)$ is given by the Werthamer-Helfand-Hohenberg theory [3].

Furthermore, the superconducting fluctuation can decrease the density of states (DOS). In our simulation, we consider that the influence on the DOS can be represented by a thermal excitation of quasiparticle. Thus, the total conductivity can be written as:

$$\delta\sigma = \sigma_0 + \frac{e^2}{\pi^2 \hbar}[\alpha I_\alpha(b,t) + \beta I_\beta(b,t)] + C\left[\exp\left(-\frac{\Delta}{k_B T}\right) - 1\right] \quad (4)$$

where $\sigma_0$ is background conductivity and $\Delta$ is the activation energy of thermal excitation (the local superconducting pairing strength). Then we try to reproduce the experimental $R_s(T)$ curves using equation (4). However, we find that the theoretical formula is in good agreement with the experimental curves only when $I_\alpha(b,t)$ dominates the conductivity correction of superconducting fluctuation and $I_\beta(b,t)$ term is small. Finally, we set $\beta = 0$ and estimate $\alpha$ about 1/3 according to the amplitude of the reentrant feature at low temperatures. The $I_\beta(b,t)$ term can be neglected when approaching low temperature limit [1,2,4]. Therefore, the total conductivity reads:

$$\delta\sigma = \sigma_0 + \frac{e^2}{3\pi^2\hbar}\left[\ln\frac{r}{b} - \frac{1}{2r} - \psi(r)\right] + C\left[\exp\left(-\frac{\Delta}{k_B T}\right) - 1\right] \quad (5)$$

The minimum and maximum ($dR/dT = 0$) of conductivity correction (equation (5)) from 3.5 T to 4.1 T give rise to the theoretical phase boundary since they separate the regions of $dR/dT < 0$ (weakly localized metal state) and $dR/dT > 0$ (superconducting state). By adjusting three coefficients $\sigma_0$, $C$ and $\Delta$ in the equation (5), the theoretical phase boundary is consistent with the experimental observations [Fig. 2(b) in the main text] and the reentrant feature of experimental $R_s(T)$ curves [Fig. S1] can also be reproduced. The parameters are summarized in Table S1.

## III. Figures and Tables

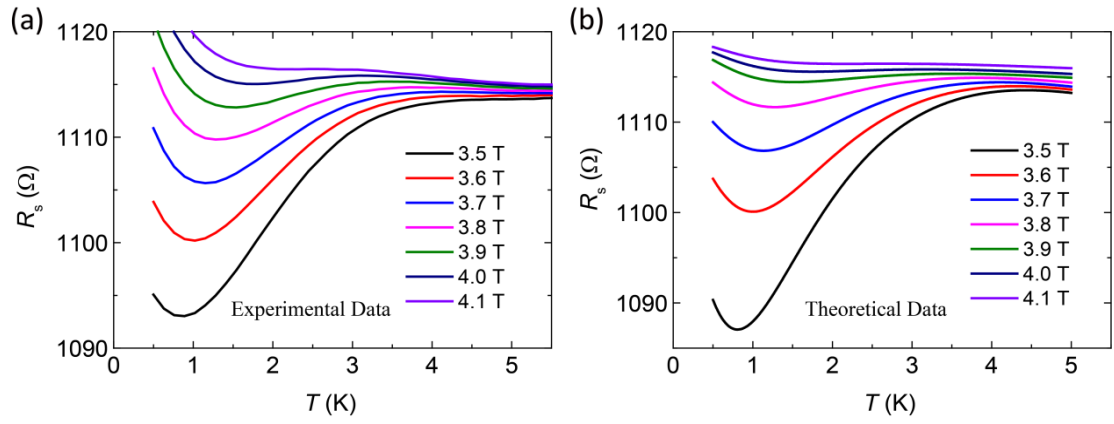

Figure S1 (a) The measured $R_s(T)$ curves of 4-ML Pb film from 3.5 T to 4.1 T. (b) Theoretical $R_s(T)$ curves at various magnetic fields considering the influence of superconducting fluctuations (see Supplemental Material Part II for details). The theoretical curves are qualitatively consistent with the experimental data of 4-ML Pb film.

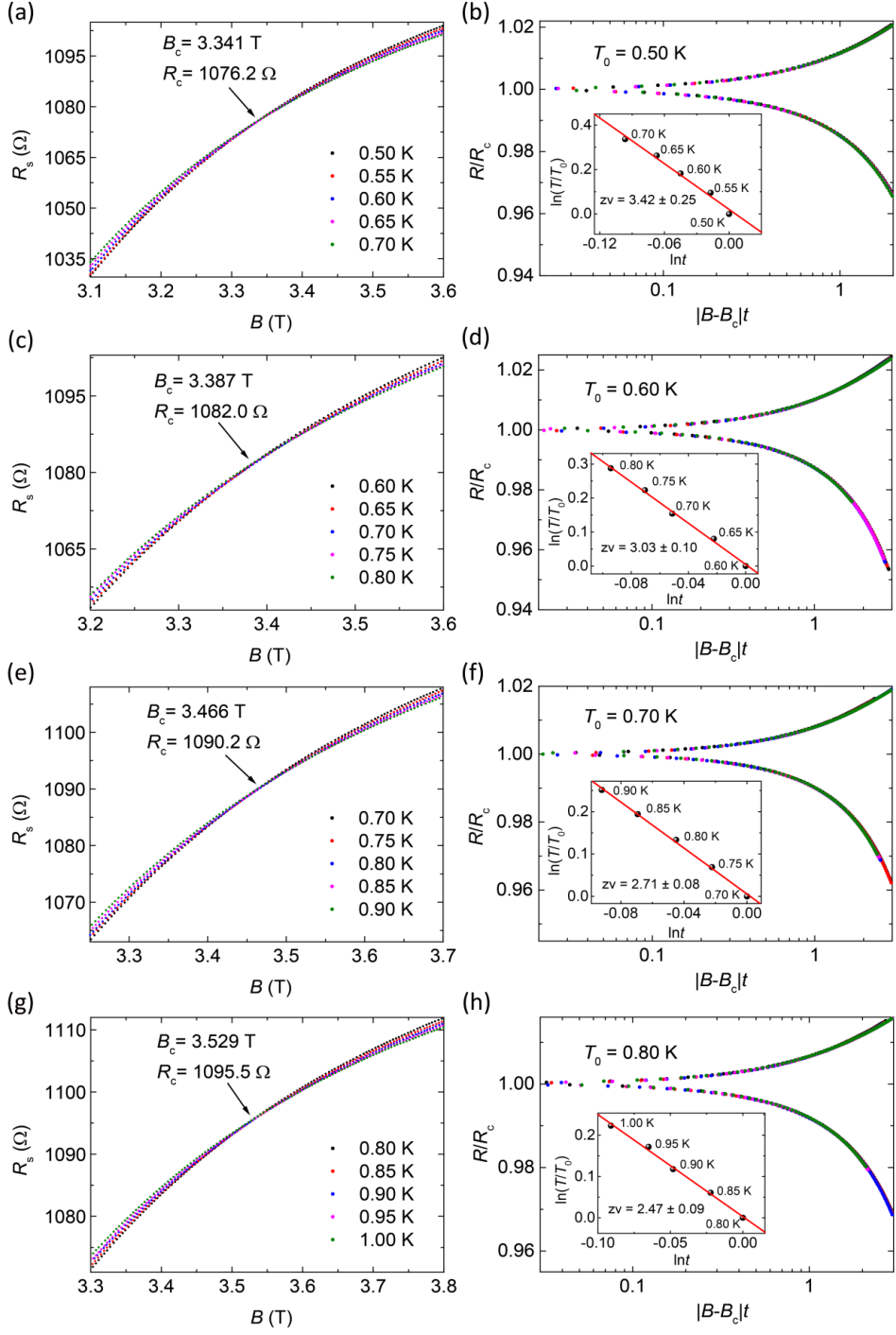

Figure S2 Finite-size scaling analysis for 4-ML Pb film at temperatures from 0.5 to 1.0 K. (a)(c)(e)(g) Sheet resistance as a function of magnetic field close to SMT boundary at various temperature ranges of (a) 0.5-0.7 K, (c) 0.6-0.8 K, (e) 0.7-0.9 K and (g) 0.8-1.0 K. (b)(d)(f)(h) Corresponding normalized sheet resistance as a function of scaling variable $|B - B_c|t$, with $t = T/T_0^{-1/zv}$. Inset: linear fitting between $\ln(T/T_0)$ and $\ln(t)$ gives critical exponent $zv$.

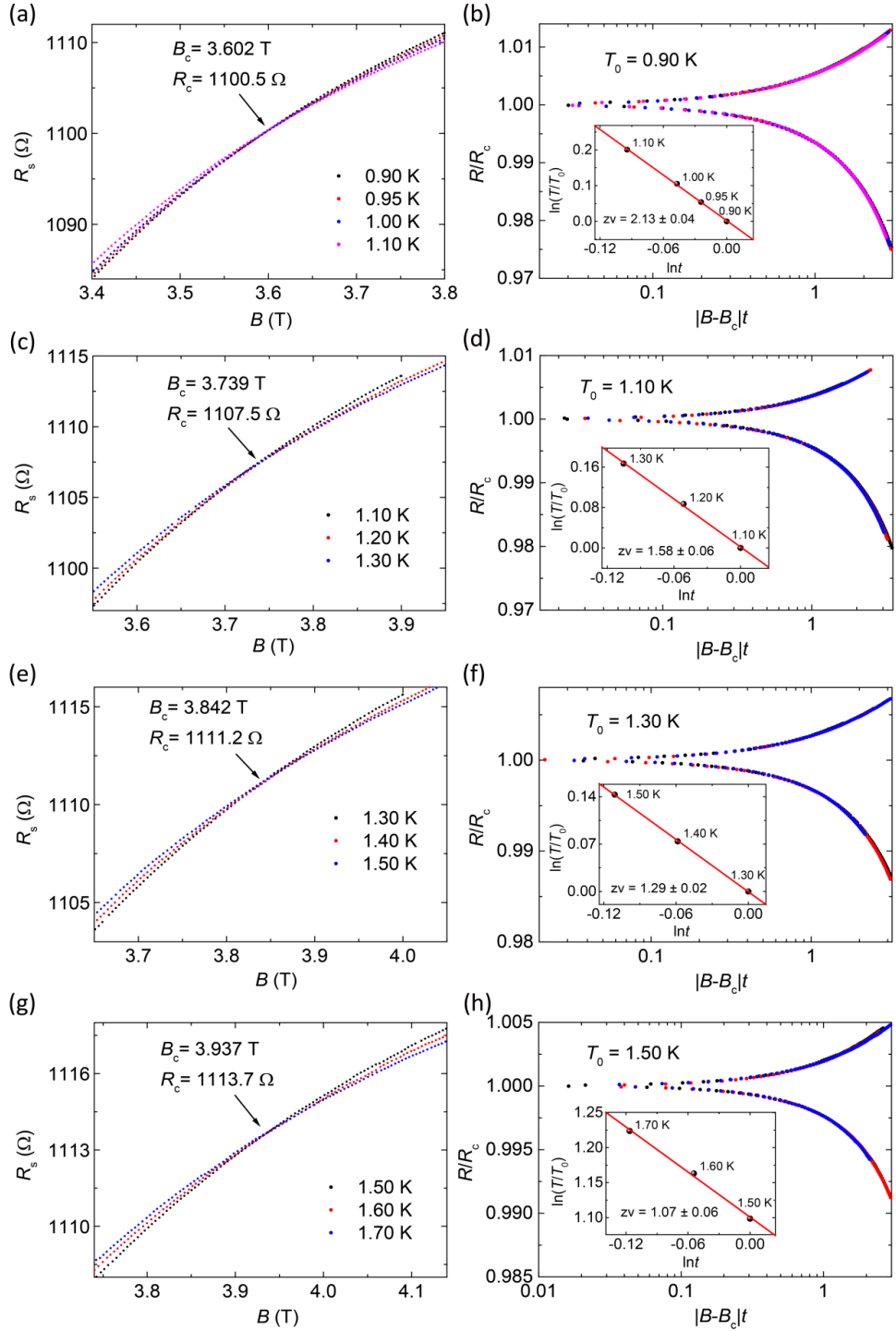

Figure S3 Finite-size scaling analysis for 4-ML Pb film at temperatures from 0.9 to 1.7 K. (a)(c)(e)(g) Sheet resistance as a function of magnetic field close to SMT boundary at various temperature ranges of (a) 0.9-1.1 K, (c) 1.1-1.3 K, (e) 1.3-1.5 K and (g) 1.5-1.7 K. (b)(d)(f)(h) Corresponding normalized sheet resistance as a function of scaling variable $|B - B_c|t$, with $t = T/T_0^{-1/zv}$. Inset: linear fitting between $\ln(T/T_0)$ and $\ln(t)$ gives critical exponent $zv$.

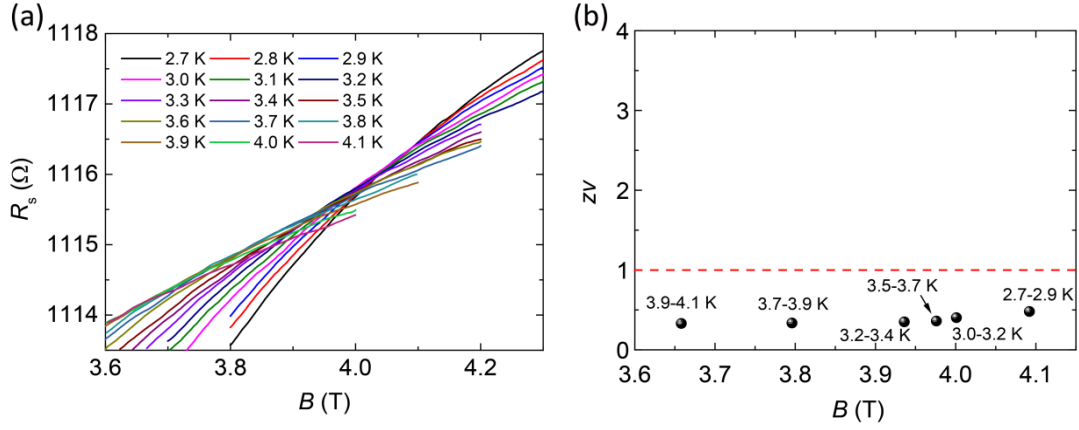

Figure S4 (a) $R_s(B)$ curves in relatively high temperature region between 2.7 K and 4.1 K in 4-ML Pb film exhibit a pronounced transition region rather than a critical point. (b) Critical exponent $zv$ from 2.7 K to 4.1 K in 4-ML Pb film. In relatively high temperature regime, the $R_s(B)$ curves also exhibit a crossing region, but zv is small (<1) and relatively stable.

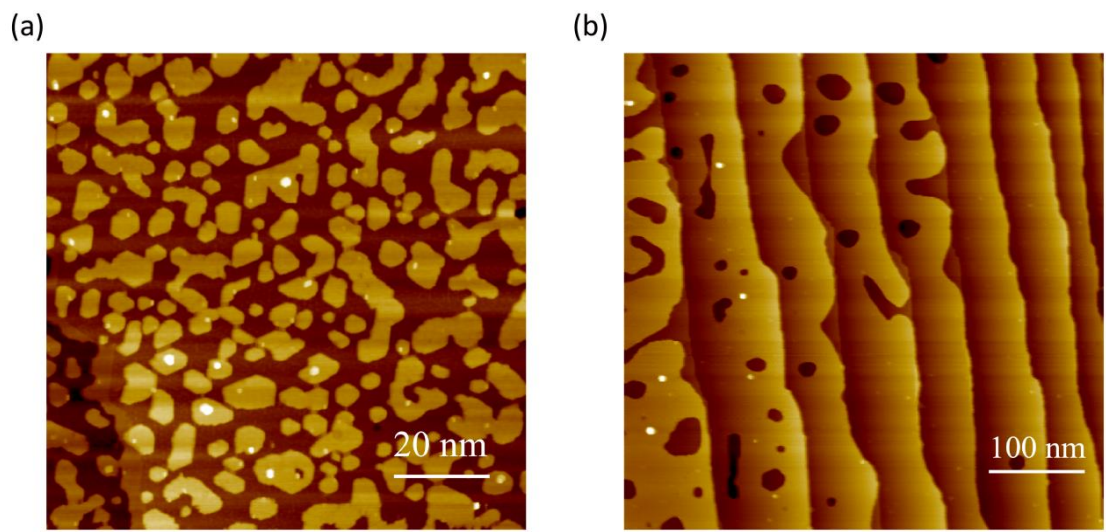

Figure S5 Typical STM images of (a) 3.5-ML and (b) 4-ML Pb film [5].

Table S1 The parameters of equation (5) ($\delta\sigma = \sigma_0 + \frac{e^2}{\pi^2\hbar}\alpha I_\alpha(b,t) + C\left[\exp\left(-\frac{\Delta}{k_B T}\right) - 1\right]$) at different magnetic fields from 3.5 T to 4.1 T.

| $B$ (T) | $\alpha$ | $C$ ($h/4e^2$) | $\Delta/k_B$ (K) | $\sigma_0$ ($h/4e^2$) |
|---|---|---|---|---|
| 3.5 | 1/3 | 0.58 | 10.4 | 6.24 |
| 3.6 | 1/3 | 0.51 | 9.9 | 6.18 |
| 3.7 | 1/3 | 0.45 | 9.2 | 6.12 |
| 3.8 | 1/3 | 0.37 | 8.0 | 6.04 |
| 3.9 | 1/3 | 0.31 | 7.2 | 5.98 |
| 4.0 | 1/3 | 0.29 | 7.3 | 5.97 |
| 4.1 | 1/3 | 0.27 | 7.4 | 5.95 |